\documentclass[aps,prd,preprint,12pt,superscriptaddress,nofootinbib,a4paper]{revtex4-1}

\usepackage[utf8]{inputenc}
\usepackage{amsmath,amssymb}
\usepackage[separate-uncertainty=true]{siunitx}
\usepackage{graphicx}
\usepackage[usenames,dvipsnames]{xcolor}
\usepackage[caption=false]{subfig}
\usepackage{hyperref}
\usepackage[capitalize]{cleveref}
\usepackage{booktabs}
\usepackage{tabularx}
\usepackage{xspace}
\usepackage{color}
\usepackage[normalem]{ulem}
\usepackage{slashed}
\usepackage{bbold}
\usepackage{wasysym}
\usepackage{graphicx}
\usepackage{mathrsfs} 
\usepackage{soul}

\allowdisplaybreaks

\graphicspath{{graphics/}}

\newcommand{\eqn}{equation}
\newcommand{\lb}{\left(}
\newcommand{\rb}{\right)}

\newcommand{\al}{\alpha}

\newcommand{\GeV}{{\ensuremath\rm GeV}}

\newcommand{\pb}{{\ensuremath\rm pb}}
\newcommand{\fb}{{\ensuremath\rm fb}}

\DeclareSIUnit{\pb}{pb}
\DeclareSIUnit{\fb}{fb}
\AtBeginDocument{
\heavyrulewidth=.08em
\lightrulewidth=.05em
\cmidrulewidth=.03em
\belowrulesep=.65ex
\belowbottomsep=0pt
\aboverulesep=.4ex
\abovetopsep=0pt
\cmidrulesep=\doublerulesep
\cmidrulekern=.5em
\defaultaddspace=.5em

\newcolumntype{C}{>{\centering\arraybackslash}X}
\newcolumntype{b}{C}
\newcolumntype{s}{>{\hsize=.6\hsize}C}
\newcolumntype{R}{>{\raggedleft\arraybackslash}X}
}

\begin{document}

\bibliographystyle{hunsrt}
\date{\today}
\rightline{RBI-ThPhys-2022-20}
\title{{\Large Signal strength and W-boson mass measurements as a probe of the electro-weak phase transition at colliders - Snowmass White Paper}}

\author{Andreas Papaefstathiou}
\email{apapaefs@cern.ch}
\affiliation{Department of Physics, Kennesaw State University, Kennesaw, GA 30144, USA}

\author{Tania Robens}
\email{trobens@irb.hr}
\affiliation{Ruder Boskovic Institute, Bijenicka cesta 54, 10000 Zagreb, Croatia}

\author{Graham White}
\email{graham.white@ipmu.jp}
\affiliation{Kavli IPMU (WPI), UTIAS, The University of Tokyo, Kashiwa, Chiba 277-8583, Japan.}

\renewcommand{\abstractname}{\texorpdfstring{\vspace{0.5cm}}{} Abstract}

\begin{abstract}
    \vspace{0.5cm}
    We consider an extension of the scalar
  sector of the Standard Model (SM) by an additional gauge singlet, which mixes with a
  part of the SM-like Higgs doublet. Within this model,
  parameter-space regions exist that can lead to a strong first-order electro-weak
  phase transition, a necessary condition for electro-weak
  baryogenesis. We discuss how such regions of the parameter space can
  be tested using the SM-like Higgs boson's signal strength measurements, as well as precision observables, such as e.g.\ the W-boson mass, at current and future colliders.
\end{abstract}

\maketitle
\section{Introduction}
In this short white paper, we discuss a specific new physics scenario
that enables a strong first-order electro-weak phase transition
(SFO-EWPT).  If cosmological electro-weak symmetry breaking occurred
through a SFO-EWPT, it could provide an explanation as to why there is
so much more matter than anti-matter in the
universe, see e.g.~\cite{Morrissey:2012db,White:2016nbo}. The model on which we
focus on extends the SM scalar sector by an additional
scalar field that does not possess any SM charges, i.e.\ a gauge
singlet. The model allows for mixing between the two fields, as well
as novel scalar-scalar interactions. We build this white paper upon
the discussions of refs.~\cite{Papaefstathiou:2020iag,Papaefstathiou:2021glr}. Similar studies, which however partially focus on regions that are mainly sensitive to $h_{125}\,\rightarrow\,h_j\,h_j$, have e.g. been presented in \cite{Chen:2017qcz,Kozaczuk:2019pet,Carena:2019une,Carena:2022yvx,Liu:2022nvk}.

\section{The model}
We consider a new physics scenario where the SM scalar sector is
supplemented by an additional real gauge-singlet field. Such scenarios
can be realized with or without imposing additional symmetries on the
potential. We work here in the most general scenario, where we assume
renormalizability of the model. The potential that includes the
additional terms in the gauge eigenbasis is then given by
\begin{eqnarray}\label{eq:xsm}
V(H,S) &=& \mu^2 (H^\dagger H) + \frac{1}{2} \lambda (H^\dagger H)^2 + K_1 (H^\dagger H) S \\ \nonumber
             &+& \frac{K_2}{2} (H^\dagger H) S^2 + \frac{M_S^2}{2} S^2 + \frac{\kappa}{3} S^3 + \frac{\lambda_S}{2} S^4 \;,
\end{eqnarray}
where $S$ denotes the new gauge-singlet eigenstate and $H$ is the Higgs doublet. Note that this model has in total five new parameters with respect to the SM, represented here by the variables $K_1$, $K_2$, $M_S$, $\kappa$ and $\lambda_S$.

In general, the discussion of this model and constraints is involved,
see
e.g. \cite{Chen:2014ask,Chen:2017qcz,Lewis:2017dme,Papaefstathiou:2020iag,Papaefstathiou:2021glr}. In
this work, however, we are only interested in the investigation of
constraints at current and future colliders that depend on the mass $m_H$ of
the second scalar and the mixing angle $\theta$ that connects gauge and mass
eigenstates in the scalar sector. We use the nomenclature in which $\sin\theta\,=\,0$ corresponds to the SM decoupling. The other free parameters do not play a role in the above phenomenological constraints, but are relevant for the perturbativity of the model at higher scales. In what follows, during our phenomenological scans over the parameter space of the model, we label points that become non-perturbative at 10 TeV.

\section{A strong first-order electro-weak phase transition}
Describing the nature of the electro-weak transition is an ongoing theoretical challenge~\cite{Croon:2020cgk,Niemi:2021qvp,Gould:2021oba, Niemi:2020hto}. Approximate methods can be used to estimate the nature of the electro-weak phase transition in a large multi-parameter scan. At present, in doing so, one must make a choice between gauge-dependent methods, or a gauge-independent method that does not include a resummation of divergent infrared modes at leading order~\cite{Patel:2011th}. However, enormous unphysical scale dependence was found in multiple studies, originating in the poor convergence of perturbation theory~\cite{Gould:2021oba}. This suggests that neglecting certain resummation terms results in large uncertainties. For a scalar singlet, unlike the SM, the new contributions, either to a tree level barrier or the thermal barrier, are gauge independent, due to the gauge-singlet nature of the new field. We therefore follow the methods of ref.~\cite{Papaefstathiou:2020iag,Papaefstathiou:2021glr} in using a gauge-dependent method, with leading-order ``Arnold-Espinosa'' resummation~\cite{Arnold:1992rz,Arnold:1992fb}. Specifically, we include the one-loop corrections at finite temperature, evaluated in the covariant gauge using the $\overline{\mathrm{MS}}$ scheme, and include a leading-order resummation of Daisy diagrams:
\begin{equation}
    V(h,s , T, \mu , \xi_W,\xi_Z)= V_{\rm Tree}(h,s , \mu) + V_{\rm CW} (h,s, \mu , \xi_W, \xi _Z) + V_T(h,s,T, \mu , \xi_W, \xi _Z)
\end{equation}
where $h$ and $s$ are the scalar fields after symmetry breaking, $T$ is the temperature, $\xi_{W,Z}$ are the gauge parameters, $\mu$ is the renormalization scale, $V_{\rm CW}$ is the zero-temperature one-loop Coleman-Weinberg correction and $V_T$ is the thermal potential. For details see ref.~\cite{Papaefstathiou:2020iag}.

Across the parameter space, it was found in~\cite{Papaefstathiou:2020iag} that the majority of points were only predicting a strong first-order electro-weak transition for some particular values of the unphysical renormalization scale. The points were categorized in terms of how robust the claim that the point predicts a strong first-order transition is. Here we focus on points that possess the lowest degree of theoretical uncertainty, which were dubbed ``(ultra-) conservative'' in~\cite{Papaefstathiou:2020iag}, where the phase transition was found to be strongly first order, independently of the gauge parameters and the renormalization scale, and that reproduce zero-temperature observables for \textit{some} values of the scale.\footnote{In effect, we include the ``ultra-conservative'' and ``conservative'' categories of ref.~\cite{Papaefstathiou:2020iag}.} To derive parameter-space points that satisfy SFO-EWPT, we perform random scans over the five free parameters in the potential. We generate points with viable zero-temperature phenomenology, leaving a large leeway for theoretical uncertainties. Table~\ref{tab:parameters} presents our scan range. For further detail on the approach, see ref.~\cite{Papaefstathiou:2020iag}.

\begin{table}[h]
    \centering
    \begin{tabular}{c|c}
        Parameter & Range  \\ \hline 
        $\lambda _s$ & $[10^{-4},3]$ \\
                $K_1$ & $[-700,0]$ GeV \\
                $K_2$ & $[0,8]$ \\
            $M_S^2$ & $[-800^2,800^2]$ GeV \\
        $\kappa$ & $-[1800,1800]$ GeV \\
    \end{tabular}
    \caption{The range of parameters used in our scans of the parameter space of the real singlet scalar extension of the SM.}
    \label{tab:parameters}
\end{table}

In summary, viable points were considered to be those that satisfy at least the ``(ultra-) conservative'' criteria for SFO-EWPT described above, and that satisfy current constraints coming from heavy Higgs boson searches, imposed via the \texttt{HiggsBounds} (v5.10.2)~\cite{Bechtle:2008jh,Bechtle:2011sb,Bechtle:2012lvg,Bechtle:2015pma} package. The most constraining processes applied through the \texttt{HiggsBounds} package are those of the experimental searches for heavy Higgs bosons in $pp \rightarrow VV$, i.e. vector boson pair production, conducted by the ATLAS and CMS collaborations, that appear in refs.~\cite{ATLAS:2018sbw,ATLAS:2017tlw,CMS:2018amk}. As a first-order phase transition generally requires large couplings to the Higgs boson, many points quickly become non-perturbative at higher scales. We therefore mark all points where any dimensionless coupling  grows to larger than $4 \pi$ before a scale of $10$ TeV (blue circles in figure~\ref{fig:}).

The Standard Model becomes unstable above $\mathcal{O}\lb 10^{11}\rb$~\GeV{} (see, e.g. \cite{Degrassi:2012ry,Alekhin:2012py,Buttazzo:2013uya,strumia}). The model becomes stable if the scale of instability is only a few orders of magnitude higher. This can be achieved by even a small positive correction to the $\beta$ function of the Higgs quartic coupling which is provided by $K_2$ (see, e.g.~ref.~\cite{Staub:2008uz,Papaefstathiou:2020iag}). We note that in our parameter space, all points either become non-perturbative well before the instability scale, or completely stabilize the vacuum.

\section{Constraints: signal strength and $m_W$} 
\begin{center}
\begin{figure}[th!]
\begin{center}
\includegraphics[width=0.8\textwidth]{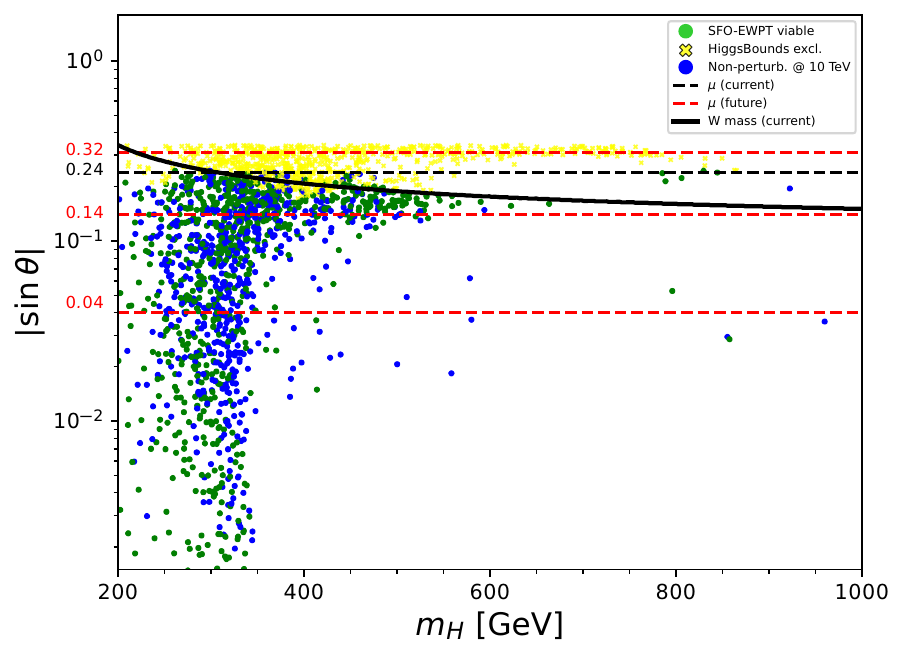}
\caption{\label{fig:} The viable points for a SFO-EWPT are shown in
  lime green filled circles. The points denoted by yellow crosses yield the necessary conditions for a SFO-EWPT, but are excluded
by direct searches for heavy scalars (imposed via the
\texttt{HiggsBounds} package). The points denoted by blue circles are
those allowed by direct searches for heavy scalars but that become non-perturbative at 10~TeV. The current W-boson mass constraint is
shown in solid black and the constraints due to signal strength
measurements of the SM-like Higgs are shown in dashed lines, with
black indicating the current constraint at 95\% C.L. ($\left| \sin \theta \right| =
0.24$) and red indicating the corresponding future constraints assuming $5\%,\,1\%,$ and
$0.1\%$ measurements with a central value of  $\mu\,=\,1$ ($\left| \sin \theta \right| =
0.32, 0.14, 0.04$, respectively). }
\end{center}
\end{figure}
\end{center}

Two constraints that only depend on the second scalar mass and the mixing angle\footnote{At the order of perturbation theory discussed here; extending to higher orders might introduce additional parameter dependencies.} are the one-loop corrections to the W-boson mass, as well as the signal strength measurements for the 125~\GeV{} scalar. The signal strength is related to the mixing angle via $\cos^2\theta\,\gtrsim\,\mu$, which, taking the current ATLAS combination value \cite{ATLAS-CONF-2021-053}, $\mu\,=\,1.06\,\pm\,0.06$, leads to $|\sin\theta|\,\lesssim\,0.24$ at 95\% C.L.. We display this bound as ``current bound" in figure \ref{fig:}. Furthermore, the precision that might be achievable at future colliders for the signal strength varies from collider to collider, reaching per-mille level at future machines \cite{caterinatalk}. We therefore also display lines, at 95\% C.L., for an assumed precision of $5\%,\,1\%,$ and $0.1\%$ in figure \ref{fig:}. For these, set $\mu\,=\,1$. Note that direct searches also constrain the parameter space; however, these depend on additional parameters, as the new physics decays influence branching ratios and in turn rates for SM-like final states.\footnote{An example can be found e.g. in~\cite{Robens:2022oue}.} Here we only impose \textit{current} constraints through the \texttt{HiggsBounds} package. 

For the calculation of the contributions to the W-boson mass, we essentially follow the work presented in \cite{Lopez-Val:2014jva}. However, since then many of the parameters used in the evaluation of the SM-like contribution have been updated, we have re-calculated the SM-prediction~\cite{Awramik:2003rn} using the most recent electro-weak parameters~\cite{Keshavarzi:2019abf,ParticleDataGroup:2020ssz}. In particular, we use the set:

\begin{eqnarray*}
\al_s\lb m_Z\rb\,=\,0.1179;&m_h\,=\,125.25\,\GeV;&m_t\,=\,172.76\,\GeV;\\
m_Z\,=\,91.1876;&\Delta \al_\text{had}\,=\,276\,\times\,10^{-4};&\Delta \al_\text{lep}\,=\,314.979\,\times\,10^{-4} \;,
\end{eqnarray*}
which gives $m_W^\text{SM}\,=\,80.356\,\GeV$ as the SM prediction. We compare to the experimental value\footnote{The new CDF result, $m_W\,=\,80.4335\,\pm\,0.0094\,\GeV$ \cite{CDF:2022hxs}, would lead to further tension between theoretical prediction and experimental value, ruling out the scenario where the second scalar boson is heavier than 125 \GeV. In addition, the singlet scenario would not be able to explain difference in mass measurements between $p\,\bar{p}$ on the one and $p\,p,\,e^+\,e^-$ colliders on the other hand.}
\begin{\eqn*}
m_W^\text{exp}\,=\,80.379\,\pm\,0.012\,\GeV.
\end{\eqn*}
The difference between theoretical and experimental value currently corresponds to a $\sim\,1.9\,\sigma$ discrepancy. One-loop corrections with a heavy second scalar lead to negative additional contributions; therefore, the W-boson mass can be used as a single strong constraint on this model. We display the constraint on the mixing angle in figure \ref{fig:} (solid black line).

\section{Summary}
We have discussed the parameter space of a simple model, where the SM scalar sector is supplemented by an additional scalar that is a singlet under the SM gauge group, and mixes with the remnant scalar field coming from the SM-like doublet. In certain regions of parameter space, this model allows for a strong first-order electroweak phase transition. We have demonstrated which regions of parameter space are currently available within this model, and how these could be further constrained from signal strength measurements, as well as with comparison to the W-boson mass as a precision observable. We find that both are important tests of the models parameter space, and a possible discovery or exclusion of such regions could certainly render insight in the evolution of the universe.
\section*{Acknowledgements}
The authors want to thank Sally Dawson for encouraging this white paper. TR thanks the CERN theory group for their hospitality while this work was completed.
\bibliography{lit}
\end{document}